\documentclass[12pt]{article}
\usepackage{epsfig}
\usepackage{hepnames}
\usepackage{units}
\newcommand\sqrts{\ensuremath{\sqrt{s}}}
\usepackage[
    doi=true,
    eprint=true,
    sorting=none
]{biblatex}
\def\Title#1{\begin{center} {\Large {\bf #1} } \end{center}}

\begin{document}

\Title{Precision Higgs Measurements at the 250 GeV ILC}

\bigskip\bigskip

\begin{raggedright}

{\it Jan Strube\index{Strube, J.}\\
Pacific Northwest National Laboratory\\
902 Battelle Blvd, Richland, WA 99352\\
\bigskip
Center for High Energy Physics\\
University of Oregon\\
1274 University of Oregon
Eugene, Oregon 97403-1274}\bigskip\bigskip
\end{raggedright}

\noindent
“Talk presented at the APS Division of Particles and Fields Meeting (DPF 2017), July 31-August 4, 2017, Fermilab. C170731”

\section{Introduction}

The International Linear Collider (ILC) is a proposed 31 km electron -- positron collider with a baseline energy of 500 GeV, with a physics program that is complementary to the LHC. In addition to precision measurements in the Higgs, top, and electroweak sectors that will deepen our understanding of electroweak symmetry breaking, the machine offers a rich discovery potential for new particles~\cite{Fujii:2017ekh}. Extensibility being a key feature of a linear collider, the design~\cite{Adolphsen:2013kya} accommodates an extension to 1 TeV. This would allow for a higher precision measurements of the Higgs self-coupling and additional reach for searches, and could be motivated by discoveries at the earlier stages of the ILC, the LHC, or other experiments. Recently, the possibility to operate the machine in stages, with a first stage at 250 GeV has been discussed at the ``Americas Workshop on Linear Colliders 2017'' at SLAC~\cite{awlc:slac:2017}. The Higgs physics program of the 500 GeV stage has been presented at a previous meeting in this conference series~\cite{Strube:2015ohu}. The purpose of this note is to present an update of this physics program for the 250 GeV stage.

\section{The Accelerator}
\begin{figure}
    \centering
    \includegraphics[width=0.49\linewidth]{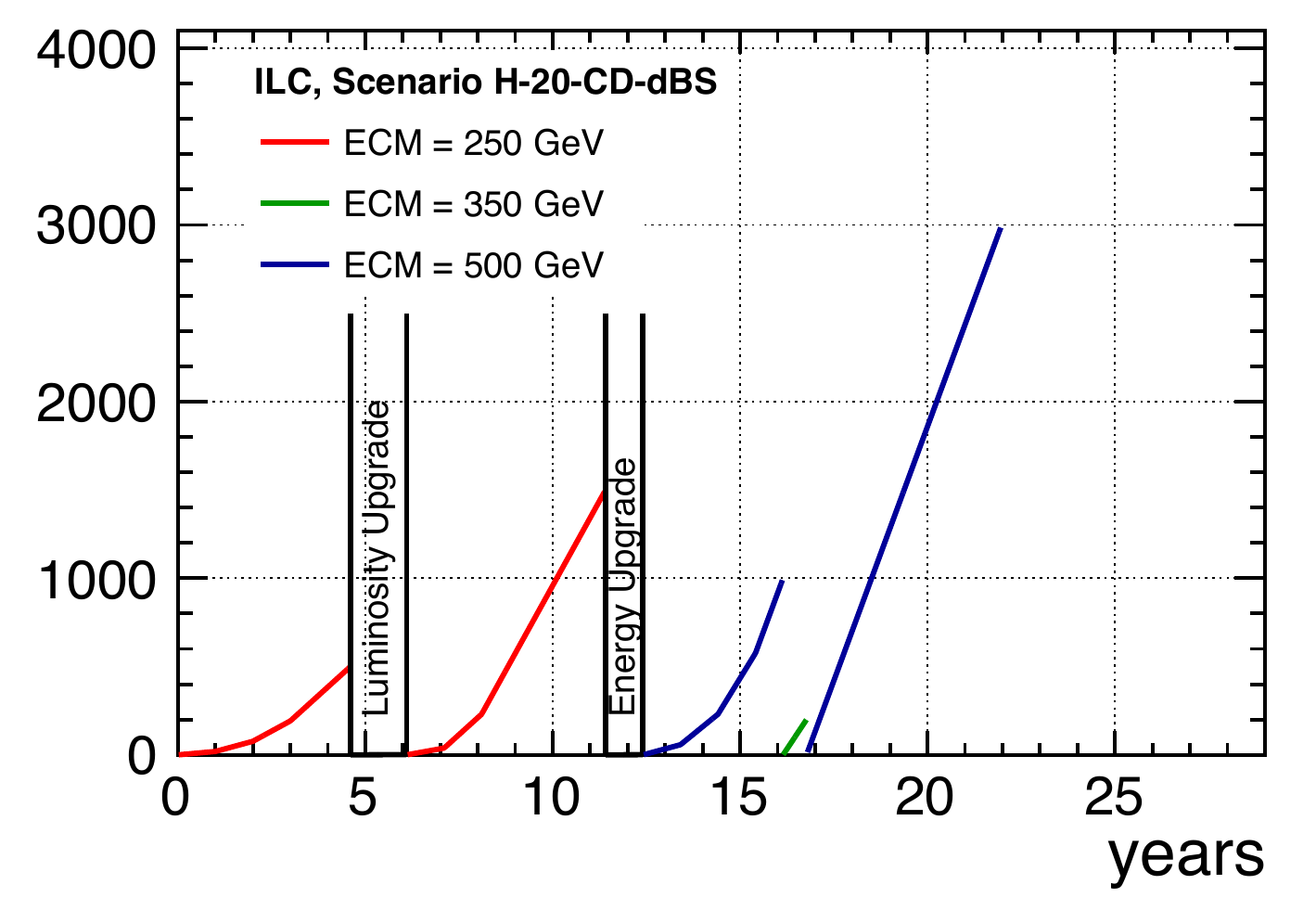}
    \caption{Candidate scenario for operating the ILC in the the first stage and in stages with higher luminosity and energy.}
    \label{fig:operatingScenario}
\end{figure}

The accelerating technology of the ILC is based on superconducting RF cavities, essentially the same technology that is being used in the free electron lasers XFEL at DESY and LCLS-II at SLAC, but with a slightly higher accelerating gradient of 31.5 MV/m.
A candidate site for the machine has been identified in the Kitakami region in the Iwate prefecture in northern Japan.
The timelines for different deliverables of the physics program in a number of candidate operating scenarios have been evaluated to guarantee the highest possible impact at the earliest time. One of the candidate scenarios for operating the machine is shown in Figure~\ref{fig:operatingScenario}.

\section{Detectors at the ILC}
Detector concepts that can deliver the required precision have been studied in two study groups for the International Large Detector (ILD) and the SiliconDetector (SiD), respectively. They are both designed with the particle flow paradigm in mind, with highly granular calorimeters that are contained in the solenoid. They feature high precision silicon vertex detectors with a trigger-less readout, a low-mass tracking system and nearly $4\pi$ coverage in solid angle. ILD features a gaseous TPC for the central tracking detector, and has a larger calorimeter system, in a 3.5 T field, while SiD is a more compact system with silicon tracking and a 5 T magnetic field. The operating scenario foresees them sharing beam time in a push--pull layout, where one of the detectors is in the beam line, while the other one is rolled out to a parking position. The performance of the baseline design for each detector concept has been studied in extensive simulation campaigns~\cite{Behnke:2013lya}. The requirement to carry out precision Higgs physics measurements has a large influence on the optimization of the detector design, and studies to improve the understanding of how detector parameters impact the physics performance continue.

\section{Higgs Physics of the 250 GeV Stage}
\begin{figure}
    \centering
    \includegraphics[width=.5\linewidth]{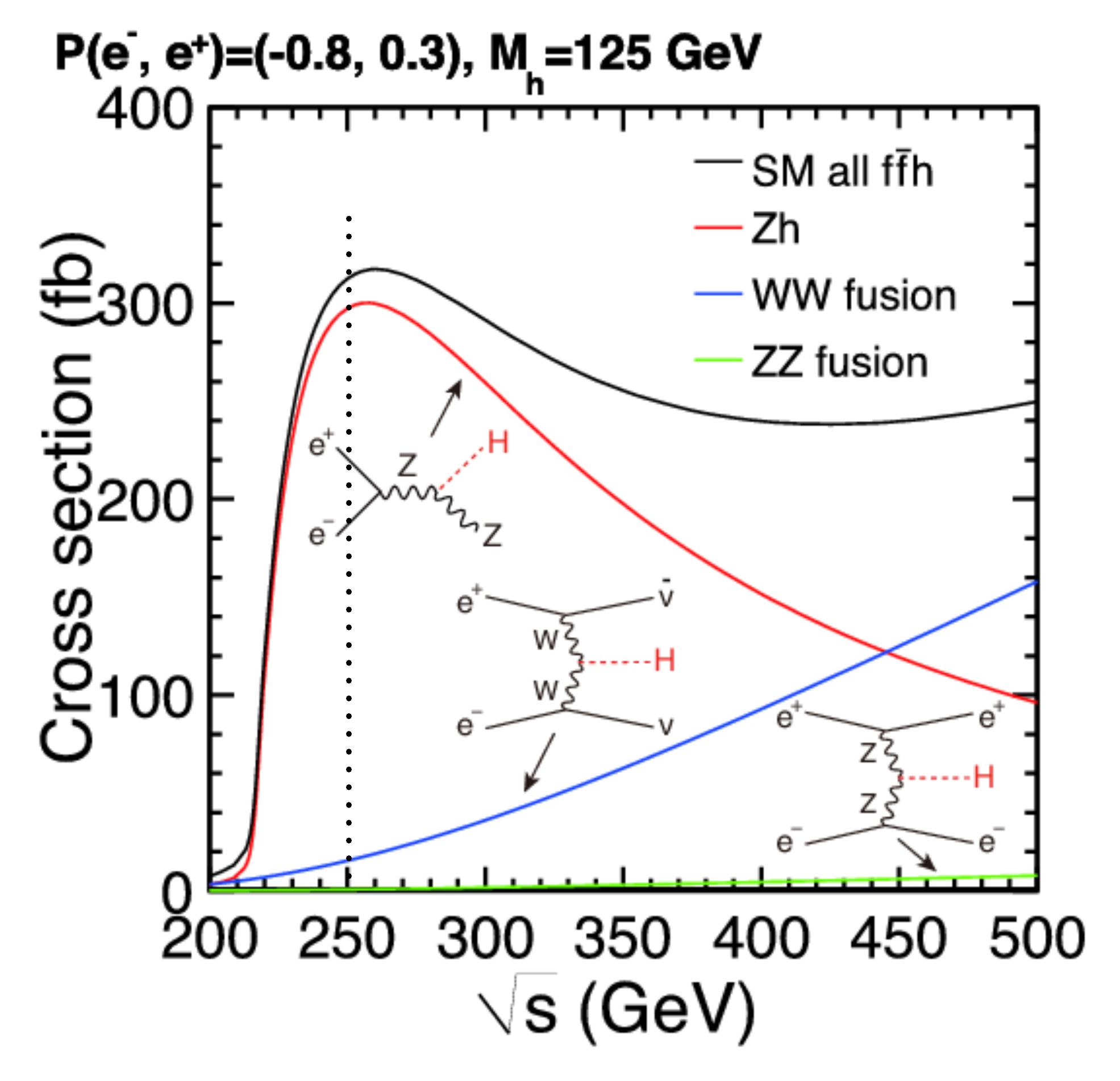}
    \caption{Standard Model cross section of the dominant Higgs production channels at different collision energies at the ILC. The polarization is assumed to be 80\% for the electron beam and 30\% for the positron beam.}
    \label{fig:higgsCrossSection}
\end{figure}

Precision measurements of Higgs properties form one of the cornerstones of the physics program at the ILC. Figure~\ref{fig:higgsCrossSection} shows the Standard Model cross sections for the dominant channels of Higgs production. At 250 GeV, the associated production with a \PZ boson is by far the largest contribution. The cross section of vector boson fusion diagrams becomes sizable at higher energies.
\subsection{The Recoil Method}
The exact determination of the Higgs production rate is one of the key measurements that enables the unprecedented precision in mapping out the Higgs sector at the ILC.
The low background and the ability to determine the collision energy at the ILC to a high degree of accuracy allows measuring the cross section of $\PZ\PH$ production independently of the $\PH$ decay in the so-called recoil technique. The $\PZ$ decay to a pair of leptons can be reconstructed very cleanly, due to the well-known mass of the $\PZ$ boson and the high momentum resolution of the tracking detectors. A fit to the recoil mass $m_{\text{rec}} = ((\sqrts - E_{\PZ})^2-p_{\PZ}^2)^{1/2}$ allows the measurement of the $\PZ\PH$ cross section without reconstructing decays of the Higgs boson. This technique is applicable at any collision energy, but it has the lowest uncertainty near threshold, where the effect of the finite spread of the collision energy due to beam--beam interaction is smallest. Recent studies have demonstrated that the bias on the cross section measurement from assumptions about the Higgs decay model can be held to less than 0.1\% in the leptonic recoil mode~\cite{Yan:2016xyx}.

\subsection{Invisible Higgs Decays}
Higgs decays to invisible final states proceed in the Standard Model through the decay $\PH\to\PZ\PZ^{*}$, where each $\PZ$ boson in turn decays to a pair of neutrinos. The branching fraction for this decay is 0.1\%~\cite{deFlorian:2016spz}. Weakly interacting massive particles, that are yet unobserved, but are hypothesized as candidates for cosmological dark matter, would lead to a signature of large missing energy in a collider detector. Higgs decays to these particles are kinematically allowed, if their mass is $m \leq m_{\PH}/2$; this would significantly alter the measured branching ratio of invisible Higgs decays.

The fact that the recoil mass technique has been shown to be independent of the Higgs decay allows the measurement of the branching fraction of unknown and hypothesized decays. This allows setting an upper limit on the branching ratio of invisible Higgs decays to 0.4\% at the 95\% confidence level~\cite{Asner:2013psa} in the ILC baseline program. This is more than an order of magnitude improvement over LHC predictions and allows for accurate test of the Standard Model prediction.

\subsection{Higgs Decays to $\tau$ Leptons}
The $\tau$ lepton is the heaviest lepton of the Standard Model and as such Higgs decays to tauons have the largest branching fraction, with the Standard Model prediction of 6.4\%. At the ILC, this would yield a sizable number of decay events that can be probed further for CP properties in an angular analysis.
Events are reconstructed in the channels $\Pq\Pq\Pgtp\Pgtm$ and $\Plp\Plm\Pgtp\Pgtm$. The analysis includes a $\tau$ jet finder that uses the jet charge to reduce background. Using the approximation that the visible tauon decay products and the neutrinos are collinear, with no other invisible decay products, the ILC program can achieve a precision on the $\tau$ Yukawa coupling of 1.9\% in the baseline and 0.9\% in a luminosity upgrade (extrapolated from the analysis at \unit[250]{GeV}~\cite{Kawada:2014gua}).

\subsection{Higgs Width}
\label{sec:higgsWidth}
An important upper bound in the measurement of the Higgs couplings, and hence the partial widths, is the total width of the Higgs boson.
The Standard Model prediction for the Higgs width is \unit[4]{MeV}~\cite{deFlorian:2016spz}, too small to be measured directly by analyzing a reconstructed mass distribution, or a production cross section threshold. In addition to serving as a constraint in a global fit, a precise measurement of this value is important for the extraction of the Higgs gauge couplings from a measurement of Higgs decays to gauge bosons, in which one of the gauge bosons is on-shell, the other is off-shell.

Under the assumption that the Higgs couplings at high invariant mass (i.e., where the Higgs is off-shell) are identical to the on-shell couplings, the HL-LHC experiments expect to be able to measure the width of the Higgs boson using decays $\PH\to\PZ\PZ^{*}$ with a precision of around \unit[22]{MeV}.
At the ILC, the Higgs width can be measured in channels where Higgs production and decay are mediated by the same coupling.
\begin{equation}
    \label{eq:definitionBR}
    \Gamma_{\PH} = \frac{\Gamma(\PH\to\PW\PW^{*})}{\mathcal{BR}(\PH\to\PW\PW^{*})}\propto\frac{g^2_{\PH\PW\PW}}{\mathcal{BR}(\PH\to\PW\PW^{*})}
\end{equation}
\begin{equation}
     \frac{g^2_{\PH\PW\PW}}{g^2_{\PH\PZ\PZ}}\propto\frac{\sigma_{\Pgn\Pgn\PH}\times\mathcal{BR}(\PH\to\Pqb\Paqb)}{\sigma_{\PZ\PH}\times\mathcal{BR}(\PH\to\Pqb\Paqb)}.
\end{equation}
At higher collision energies, the t-channel diagram for Higgs production starts to contribute more. The decay $\PH\to\PW\PW^{*}$ can be measured with greater precision than $\PH\to\PZ\PZ^{*}$, owing to the larger branching ratio. Using the definition of a branching ratio in Equation~\ref{eq:definitionBR}, one can use a high-precision measurement of a branching ratio, such as $\PH\to\Pqb\Paqb$ to substitute the coupling $g^2_{\PH\PW\PW}$ with the coupling $g^{2}_{\PH\PZ\PZ}$. The latter can be measured in the recoil analysis, independent of the Higgs boson decay.
In the full ILC program the precision on computing the Higgs width this way is 1.4\%, with the dominant contributions coming from the uncertainties on the $\PZ\PH$ cross section and the branching ratio of $\PH\to\PW\PW^{*}$.

At the 250 GeV stage, however, the t-channel diagram does not contribute much. One could instead use only the $\PZ$ -- Higgs coupling, although the branching ratio $\PH\to\PZ\PZ^{*}$ is small in the SM, about 3\%.
\begin{equation}
\frac{\sigma(\Pep\Pem\to\PZ\PH)}{\text{BR}(\PH\to\PZ\PZ^{*})} = \frac{\sigma(\Pep\Pem\to\PZ\PH)}{\Gamma(\PH\to\PZ\PZ^{*})/\Gamma_{\PH}} \propto \Gamma_{\PH}. \label{eq:hWidth}
\end{equation}

\subsection{Global Fit in the $\kappa$-Framework}
After the extraction of the full set of couplings accessible at the 250 GeV stage, the next step will be to test how consistent the Higgs boson is with the Standard Model, and more to the point, what theory best describes it. The framework that has been devised for measurements at the LHC adds deviations $\delta_{\kappa i}$ from the Standard Model couplings $\kappa_i$ as parameters in a global fit. The LHC uses seven parameters $\delta_{\kappa\PZ}$, $\delta_{\kappa\PW}$, $\delta_{\kappa\Pqb}$, $\delta_{\kappa\Pqc}$, $\delta_{\kappa\Pg}$, $\delta_{\kappa\Ptau}$, $\delta_{\kappa\Pmu}$ for the deviations from the respective Standard model couplings. In this framework, the ILC measurement of the Higgs width adds a powerful constraint.

\subsection{Effective Field theory}
Another way to understand the role of the Higgs boson is to write down an effective field theory (EFT) with all of the relevant operators.
An EFT including operators up to dimension-6~\cite{Barklow:2017awn} includes a term for the \PZ--Higgs interaction~\cite{Barklow:2017suo}:
\begin{equation}
    \delta\mathcal{L} = (1+\eta_{\PZ})\frac{m_{\PZ}^2}{v}h Z_{\mu}Z^{\mu}+\zeta_{\PZ}\frac{h}{2v}Z_{\mu\nu}Z^{\mu\nu}.
\end{equation}
A similar term can be written for the \PW--Higgs interaction. At a 250 GeV ILC, this leads to expressions for the Higgs production and decay to gauge bosons:
\begin{alignat}
        \sigma(\Pep\Pem\to\PZ\PH) &= (\text{SM}) \cdot (1+2\eta_{\PZ}&&+5.7\zeta_{\PZ}) \label{eq:eftOne}\\
        \Gamma(\PH\to\PW\PW^{*}) &= (\text{SM}) \cdot (1+2\eta_{\PW}&&-0.78\zeta_{\PW}) \label{eq:eftTwo}\\
        \Gamma(\PH\to\PZ\PZ^{*}) &= (\text{SM}) \cdot (1+2\eta_{\PZ}&&-0.50\zeta_{\PZ}) \label{eq:eftThree}.
\end{alignat}
In these expressions, the $\eta$ terms modify the SM couplings, while the $\zeta$ terms add a new type of interaction. The $\kappa$-framework does not account for these terms, and is therefore not completely model-independent. Unfortunately, these terms also introduce an asymmetry between equations~\ref{eq:eftOne} and~\ref{eq:eftThree}, so that we cannot use equation~\ref{eq:hWidth} to constrain these parameters.
Instead, one can constrain $\eta$ and $\zeta$ through angular analysis of the $\PZ\PH$ system, and through the polarization dependence of the $\PZ\PH$ channel. At the ILC, the achievable beam polarization provides a much stronger constraint than the angular analysis, and thus significantly increases the effective luminosity.

\section{Higgs Physics at later stages}
\subsection{Top Yukawa Coupling}
\begin{figure}
    \centering
    \includegraphics[width=.5\linewidth]{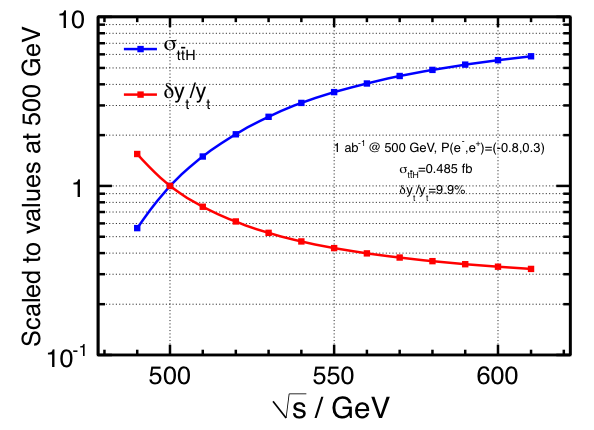}
    \caption{Cross section of the $\Pqt\Pqt\PH$ channel in fb (blue curve) and uncertainty of the top Yukawa coupling measurement for different collision energies at the ILC. The values are scaled relative to the values at 500 GeV.}
    \label{fig:topPairProduction}
\end{figure}

The top -- Higgs coupling is inaccessible at the first stage of the ILC, where the collision energy is too low to produce top quarks. A collision energy of 350 GeV opens up the top pair production channel, where the top Yukawa coupling contributes to the relevant diagrams, and can be inferred from a precision measurement of the cross section~\cite{Horiguchi:2013wra}. At higher energies, the production of the Higgs boson in association with a top quark pair allows for a direct measurement of the top Yukawa coupling. Figure~\ref{fig:topPairProduction} shows the importance of 500 GeV: The precision rapidly drops for smaller values of $\sqrt{s}$, due to the smaller cross section. Reaching a 10\% higher collision energy, on the other hand, would reduce the measurement uncertainty by about one half. In the full ILC program, including luminosity upgrades, the uncertainty on the top Yukawa coupling could reach about 3\%. A run at 1 TeV could drop this error to 2\%~\cite{Price:2014oca}.

\subsection{Higgs Self-Coupling}
The self-coupling term of the Higgs potential can be probed in multi-Higgs production at high energy colliders. The quartic coupling involving triple Higgs production is most likely inaccessible for the foreseeable future due to the small cross section of the process. Double Higgs production can be measured at the ILC in the $\PZ\PH\PH$, and, at even higher collision energy, in the $\Pgn\Pgn\PH\PH$ channels. The measurement depends on the excellent performance of b-tagging and jet energy resolution to reduce background from $\PZ\PZ\PH$ and triple $\PZ$ production. With jet clustering in addition to intrinsic detector resolution being a major source of systematic uncertainty, the current estimate for the achievable uncertainty on the tri-linear self-coupling constant in the baseline program is 27\%~\cite{Fujii:2015jha}. An energy upgrade to \unit[1]{TeV} will reduce this uncertainty to around $10\%$. The extraction of the Higgs self-coupling from an analysis of double-Higgs events is challenging, due to the many possible beyond-Standard-Model contributions. As shown by Barklow, Fujii, Jung, Peskin and Tian~\cite{Barklow:2017awn}, the precision measurements from the full ILC program will constrain the theory parameter space effectively, and therefore enable a measurement of the Higgs self-coupling with a high degree of confidence from double Higgs production.

\section{Conclusions}
\begin{figure}
    \centering
    \includegraphics[width=.5\linewidth]{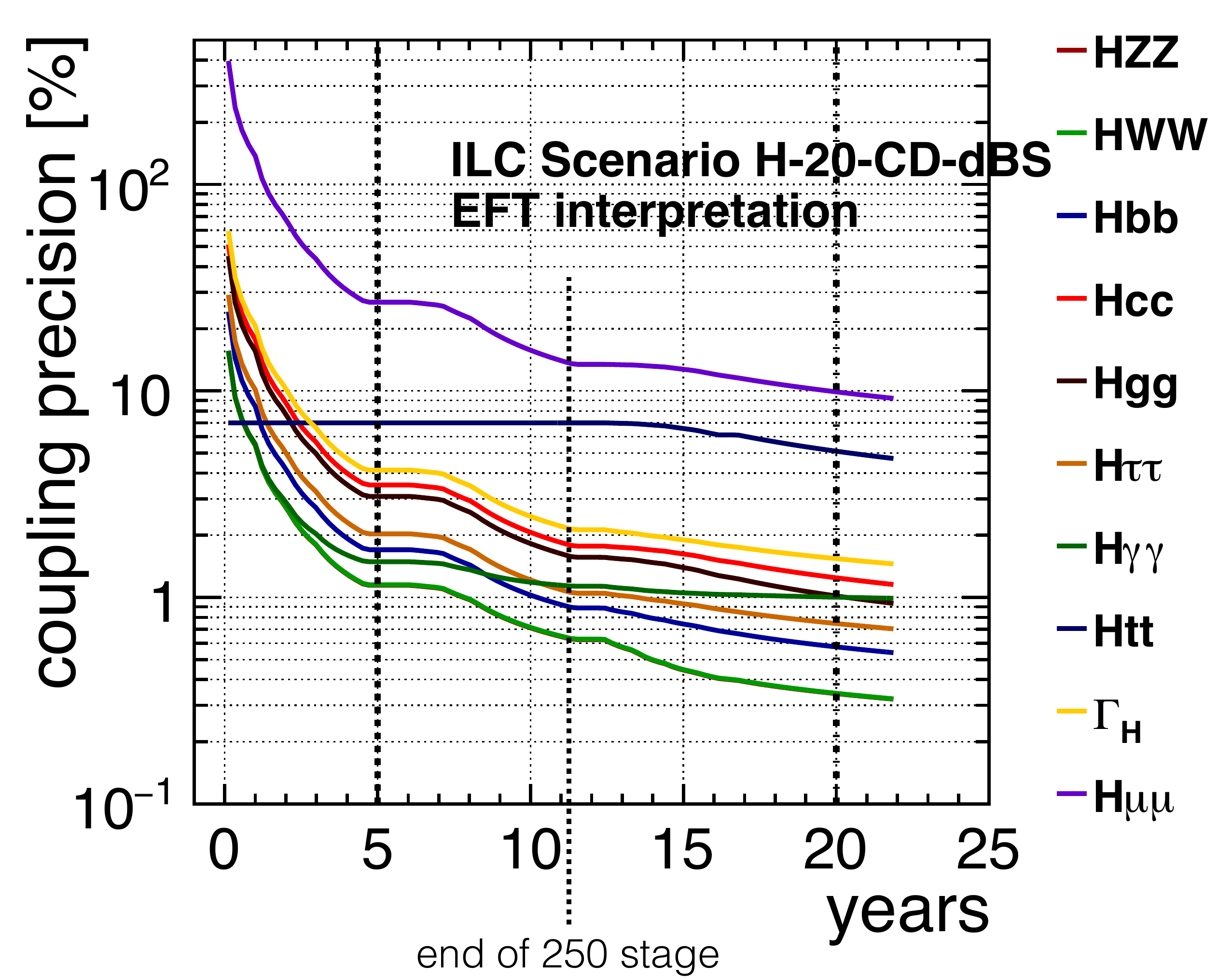}
    \caption{Evolution of the uncertainty of measurements of the Higgs width and several couplings in the ``H'' scenario. (Figure~\ref{fig:operatingScenario}).}
    \label{fig:PrecisionDevelopment}
\end{figure}

The ILC baseline program offers a comprehensive picture of the Higgs sector and allows for a self-consistent global fit of all couplings to achieve the greatest precision. Figure~\ref{fig:PrecisionDevelopment} shows the development of the coupling measurements over time in Scenario ``H''. The precision for the measurements of the properties of the Higgs boson in the full ILC program in that scenario is described in the previous sections, with additional improvements coming from runs of a \unit[1]{TeV} ILC option.

\begin{figure}
    \includegraphics[width=.49\linewidth]{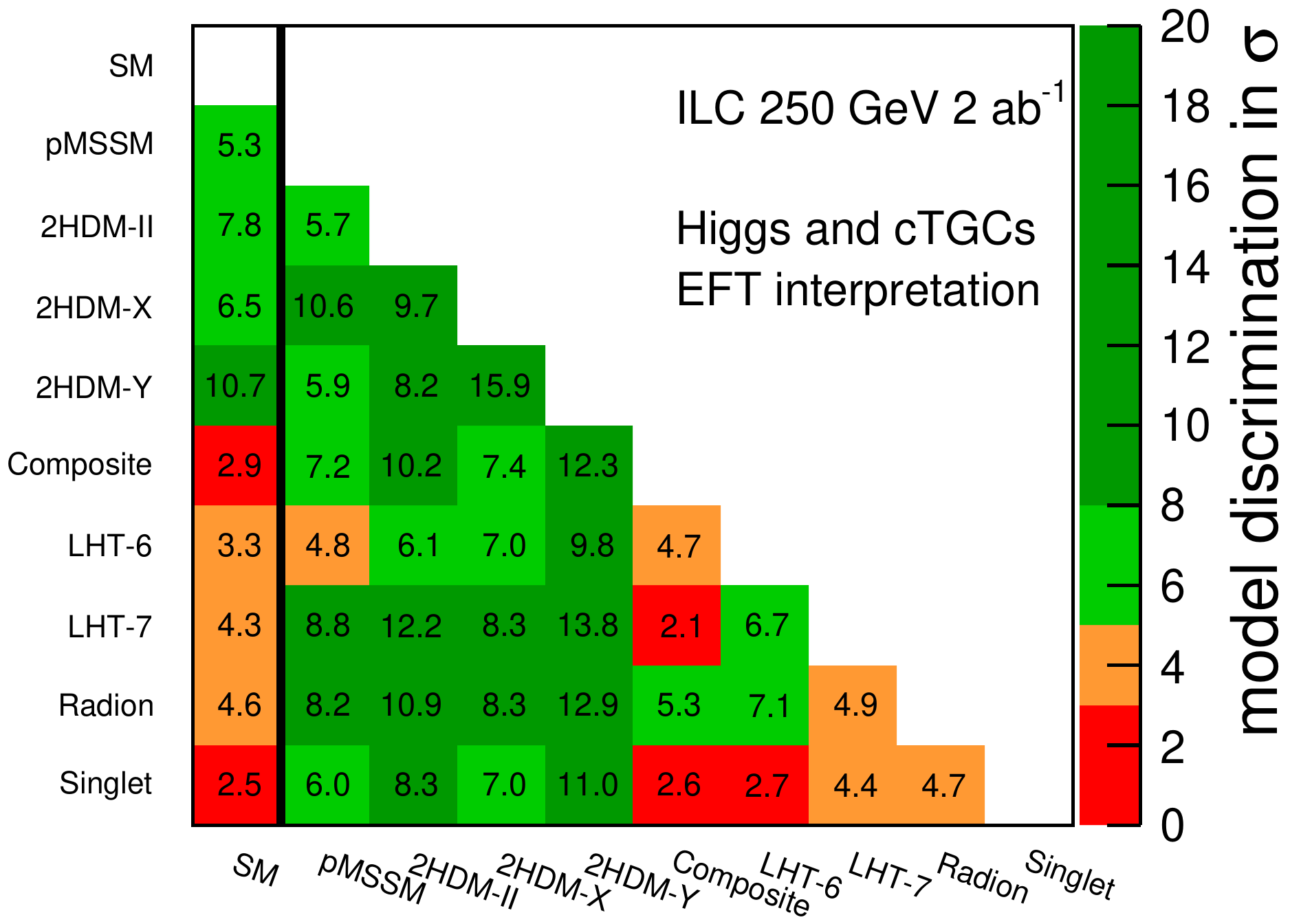}
    \includegraphics[width=.49\linewidth]{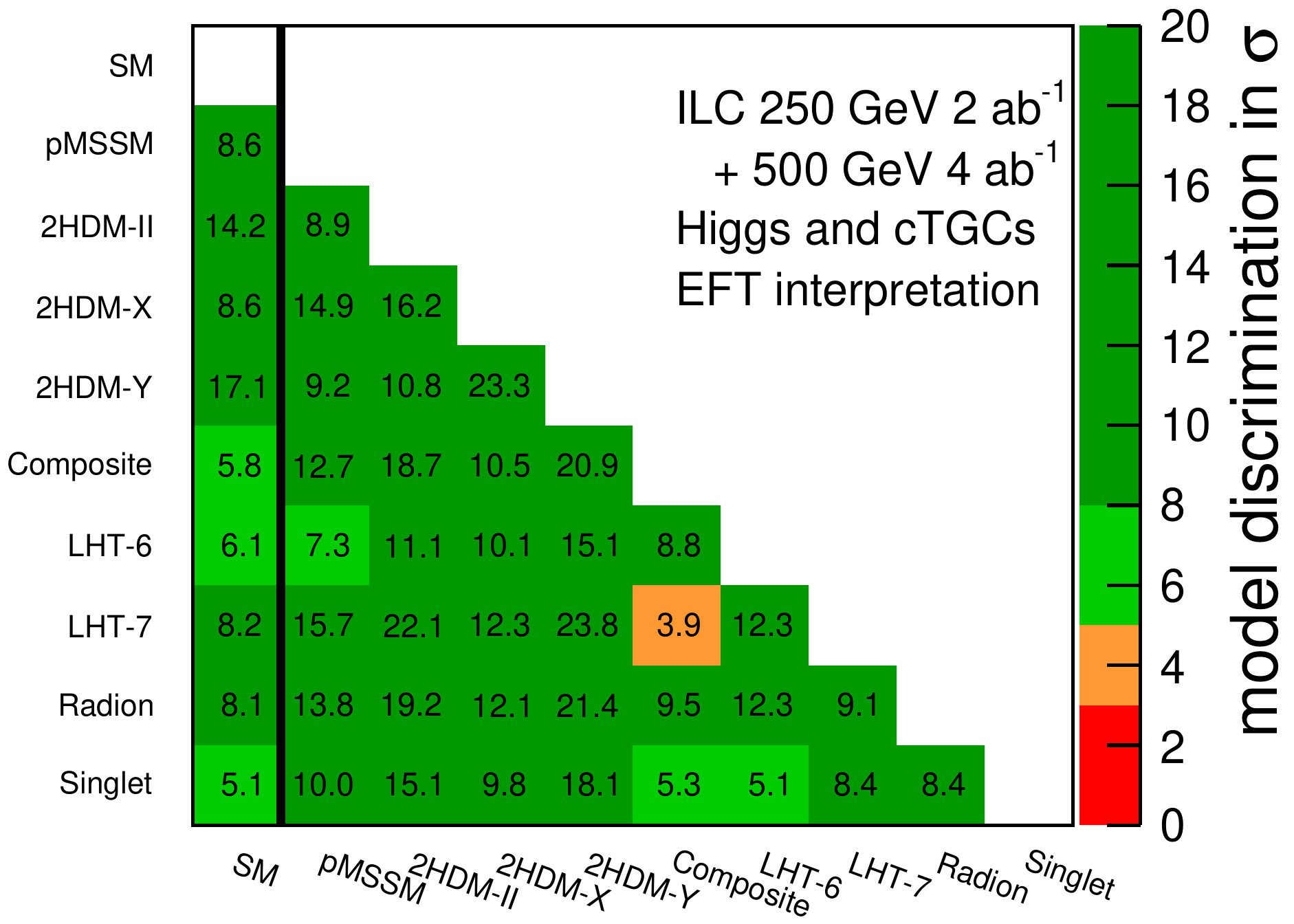}
    \caption{Graphical representation of the $\chi^{2}$-separation of the Standard Model and the several models with new physics: left: with $\unit[2]{ab^{-1}}$ of data at the ILC at 250 GeV; right with $\unit[2]{ab^{-1}}$ of data at the ILC at 250 GeV plus $\unit[4]{ab^{-1}}$ of data at the ILC at 500 GeV. Comparisons in orange have above 3 $\sigma$ separation; comparison in green have above 5 $\sigma$ separation; comparisons in dark green have above 8 $\sigma$ separation.(from~\cite{Barklow:2017suo})}
    \label{fig:modelSeparation}
\end{figure}

The constraints on new physics and potential to attribute the pattern of Higgs couplings to one of 10 select benchmark points is summarized in Figure~\ref{fig:modelSeparation}. The deviation from the Standard Model in the pattern of Higgs couplings has been simulated for a range of different models~\cite{Barklow:2017suo}. The significance (in sigma) of the separation between any two models is shown in the figure. The study shows that in the full ILC program including the 500 GeV stage, precision measurements of Higgs couplings not only allow the distinction of a given model from the Standard Model, but that in fact new physics in the Higgs sector can correctly be categorized regardless of its nature.

\clearpage

\section{Acknowledgments}
This work was presented on behalf of the global ILC detector and physics community.
\printbibliography
\end{document}